# A Low Cost ZigBee Sensor Network Architecture for Indoor Air Quality Monitoring

Tareq Alhmiedat
Department of Information Technology
Tabuk University, Tabuk, Saudi Arabia
t.alhmiedat@ut.edu.sa

Ghassan Samara
Department of Internet Technology
Zarqa University, Zarqa, Jordan
g.samara@zu.edu.jo

*Abstract*—This paper presents a low-cost system architecture that has been proposed for automatically monitoring air quality indoors and continuously in real-time. The designed system is in pilot phase where 4 sensor nodes are deployed in indoor environment, and data over 4 weeks has been collected and performance analysis and assessment are performed. Environmental data from sensor nodes are sent through ZigBee communication protocol. The proposed system is low in cost, and achieves low power consumption. Hardware and network architecture are presented in addition to real-world deployment.

## I. INTRODUCTION

Wireless Sensor Networks (WSNs) are currently an active research area due to their wide range applications including medical, environmental monitoring, safety, military, and civilian [1, 2, 3]. Habitat and environmental monitoring represent an important class of sensor network applications. Recent advances in low-power wireless network technology have created the technical conditions to build multi-functional tiny sensor devices, which can be used to sense and observe physical phenomena. Many environmental monitoring examples of WSNs are already presented in the literature and developed for different purposes.

A sensor network is a group of sensor nodes, which are low in cost and have short communication range, distributed over the area of interest. Sensor node architecture is presented in Figure 1, in which a sensor node consists of four subsystems as, are as follows: processing subunit which is responsible for processing the data captured from sensor nodes before transmitting to the base station. The sensing subunit is a device that produces a measurable response to a change in a physical condition like humidity or temperature. The communication subunit on the other hand, includes a short radio range used to communicate with neighboring nodes. And finally, the power supply subunit which contains a battery source for feeding the aforementioned units [4].

Nowadays, environment monitoring is considered as a significant area, due to the different types of pollutants spread over the world. Basically, people think that pollution is being outdoors, but the air in houses or offices could also be polluted, where the sources of indoor pollution include: tobacco smoke, gases such as carbon monoxide, and mold and pollen. Indoor Air Quality (IAQ) is the air quality within and around buildings, and structures. IAQ relates to the health and comfort of building occupants. Indoor pollution affects human health, where some health effects may show up shortly after a single exposure to pollutant. These include irritation of nose, eyes, and throat, headache, dizziness, and fatigue. Therefore, there is a great demand to monitor and control the air quality indoors.

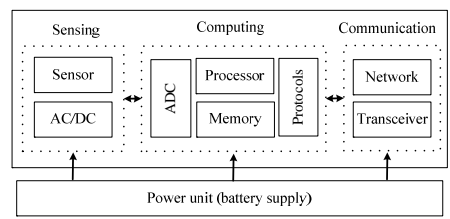

Figure 1. Sensor node architecture

In this research work, a new implementation for indoor air quality monitoring is presented, where the proposed system is designed using a low-cost architecture. The proposed air quality monitoring system integrates a single-chip microcontroller, several air pollution sensors, and a ZigBee transmission module. Authors focused on WSN data gathering methods in [5], however, in this paper, authors investigate different ways to minimize the sensor node cost and power consumption.

The remainder of this paper is organized as follows: Section 2 reviews the existing relevant works, where the system model is presented and discussed in Section 3. Section 4 presents the hardware architecture used to assess the efficiency of the proposed system. Results obtained from real experiments are discussed and analyzed in Section 5. And finally, Section 6 presents a conclusion and future works.

## II. RELATED WORKS

A wide range of recent indoor air quality monitoring systems are presented in the literature, the work presented in [6] summarizes the recently developed environment monitoring systems, which have been deployed in both environments (indoor and outdoor). This paper targets the indoor environment monitoring research area.

Various number of WSN-based indoor air quality monitoring systems have been designed and implemented recently, with different goals. A number of the existing systems target the power consumption issue [7, 8, 9, 10], others target the types of air quality sensors [9, 11, 12], and a few target the sensor node cost [13, 14]. However, in this paper, authors target several issues for indoor air quality monitoring, including: power consumption, design cost, and communication protocol.

### III. SYSTEM DESIGN

This section presents the system design architecture for indoor environment monitoring system, which is presented in Figure 2.

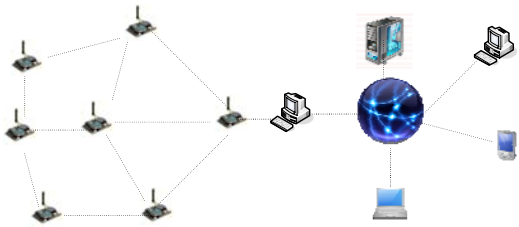

Figure 2. System architecture

The sensor node design is depicted in Figure 3, where each sensor node consists of three components, as follows:

- Array of sensors: this includes a number of sensors that attached to the sensor node, in order to measure the level of air quality indoors, temperature and humidity.

- Microcontroller: a tiny microcontroller is attached to the sensor node module, where its responsibilities include: obtain the sensors' data, handle sleep state for both sensors and communication module, and aggregate the gathered data from adjacent sensor nodes.

- Communication module: this allows the sensor node to exchange the collected data with adjacent sensor nodes in its range.

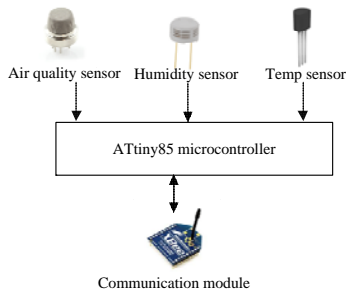

Figure 3. Sensor node design

The presented work is based on ZigBee communication protocol. ZigBee offers three main roles: coordinator, router, and end-device. A single coordinator is required for each ZigBee network, where it's role includes initiating the ZigBee network. On the other hand, router node is an optional network component, where it participates in multi-hop routing of messages. End device node is a low cost, and low power consumption sensor node, where each end-device node communicates to a router node or a coordinator node. End-device node does not participate in routing of messages and may enter sleep mode.

In this work a single coordinator and 3 router nodes have been employed in the implementation phase, however the router nodes may enter sleep mode in order to minimize the power consumption required for environment monitoring applications. Each router node is required to be awaken as soon as the collected data has been classified as significant data (data over certain threshold). In addition, router node is activated regularly to process any requests coming from sink or neighbor nodes. Figure 4 shows the flowchart for the proposed system at the sensor node.

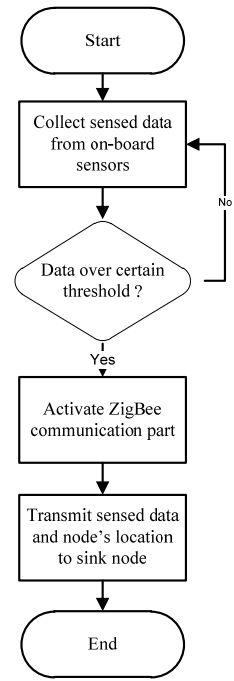

Figure 4. Flowchart for the sensor node

### IV. SYSTEM IMPLEMENTATION AND EVALUATION

In this section, the designed sensor node platform is presented, in addition to presenting the experiment test-bed. Furthermore, this section presents results obtained from several experiments which have been carried out to assess the efficiency of the proposed system.

#### A. Sensor Node Architecture

The sensor node architecture has been developed based on ZigBee protocol. The designed module consists of three sensor devices (air quality, temperature, and humidity), ATtiny85 microcontroller for processing gathered data and XBee series 2 module for communication between sensor nodes. The sensor node design is presented in Figure 5. The hardware specification for the aforementioned components is presented in Table 1.

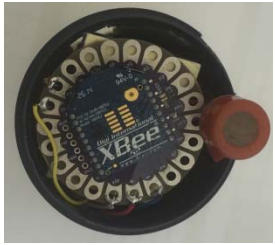

Figure 5.   The designed module with onboard sensors

TABLE I.    HARDWARE SPECIFICATION

| | Device | Manufacturer | Power consumption |
|---|---|---|---|
| Sensors | MQ-135 | Winsen Electronics Tec. | ≤ 900 mW |
| | Humidity sensor | Winsen Electronics Tec. | < 900 mW |
| | Temp. LM 36 | Sparkfun | 60 μA |
| Wireless transceive | XBee series 2 | Digi. | 40 mA |
| | XBee Pro. | Digi. | 62 mA |
| | LilyPad XBee | Sparkfun | - |
| Processor | ATtiny85, AVR 8 Pin, 20 MHz 8K, | Atmel | 300 mA |

Gas sensors are energy hungry, and hence the limited power source for sensor node will drain fast. Therefore, an interface circuit has been designed to manage the sleep state for each sensor device, in order to minimize the power consumption required for such sensors. Figure 6 depicts the experiment test-bed, which consists of 4 sensor nodes deployed indoor over an area of interest.

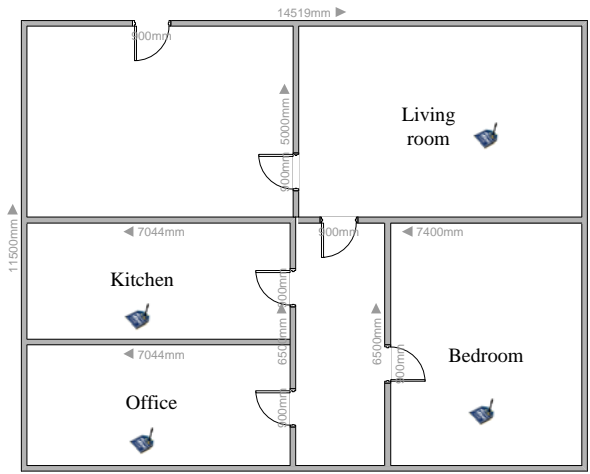

Figure 6.   Experiment test-bed

### B. System Evaluation

In this research project, a low-cost sensor node architecture has been designed and implemented, to monitor the indoor air quality, with respect to the state of the art. A number of experimental results are presented at the node and network levels to analyze the network performance.

This section includes a historical data collected from 4 stationary sensor nodes distributed indoors, for one month period (11th February to 11th March 2016). Figure 7 shows the temperature values over 24 hours' time. As presented in Figure 6, the temperature level raises up for kitchen-node, at afternoon time (13:00 – 18:00) where cooking activities are taken place.

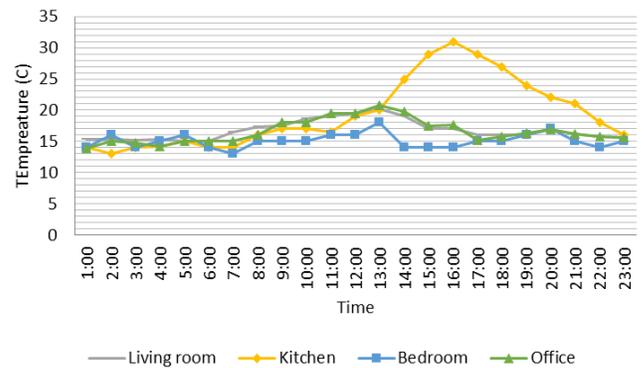

Figure 7.   Temperature values over 24 hours' time

Figure 8 shows the average air quality values collected from MQ-135 sensor, which attached to each senor node. Level of pollutant gases are increased in the kitchen-node as soon as the cooking activities are taken place, therefore the conductivity increases with air pollution sensor. MQ-135 reacts to ammonia (NH3), nitrogen oxide (NOx), benzene, and CO2 gases. Figure 9 shows the average quality values over one month period, where the kitchen node presents the high level of air quality values due to the cooking activities which are taken place over there.

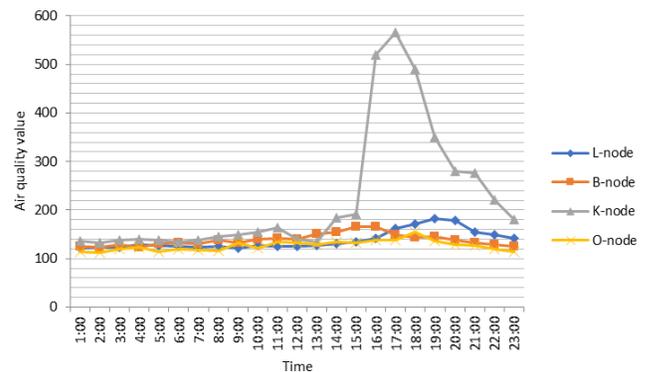

Figure 8.   Air quality values over 24 hours' time

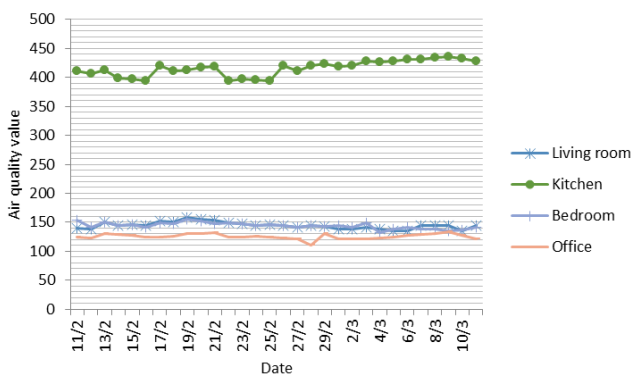

Figure 9. Average AQI values over 30 days' time

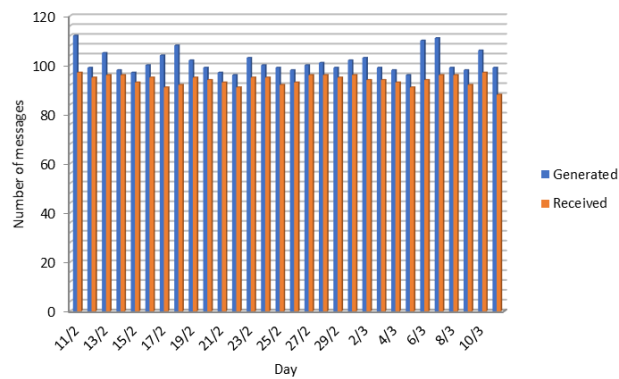

Figure 10. Throughput over 30 days' time

Next, the power consumption is evaluated. Sensor nodes are required to transmit the collected data from onboard sensors every $t$ time, where $t$ is the sampling frequency rate. An adaptive sampling frequency system is adopted, where it is based on the significance of the gathered data obtained from onboard sensors, i.e. if the gathered data is significant (high temperature, or high level of CO gas), then it is required to reach the sink node immediately. However, if the sensed data is normal data, then it will be stored on the sensor node itself, and average values of AQI, temperature, and humidity values are then transmitted to sink node every 15 minutes.

To save the limited power for sensor nodes, a sleep function is adopted, where each sensor node can enter a sleep mode for short period of time, in order to minimize the power consumption for such nodes. Transceivers modules are required to go to sleep mode whenever there is no data to send or data to receive. As soon as the sensor node is awaken, then it will check if any request coming from neighboring nodes or sink node. Otherwise, sensor node will go back to sleep mode.

Figure 10 presents the average number of messages generated at the sensor node, and received at the sink node side. The average throughput was around 0.8, this is because a small number of sensor nodes have been deployed in the experiment test-bed, moreover, the sampling frequency rate was set to reasonable value. Therefore, the aforementioned two factors have also minimized the power consumption for sensor nodes.

Kitchen node is participating in forwarding the sensed data from Living-room node to the sink node. Therefore, Kitchen-node will drain its energy first. Figure 11 presents the average power consumption for each sensor node. Office-node consumes more energy compared to other nodes, because in addition to sensing the environment, it participates in forwarding sensed data from other nodes to the sink node, whereas nodes Living room-node, Bedroom-node, and Kitchen-node consume the same amount of energy.

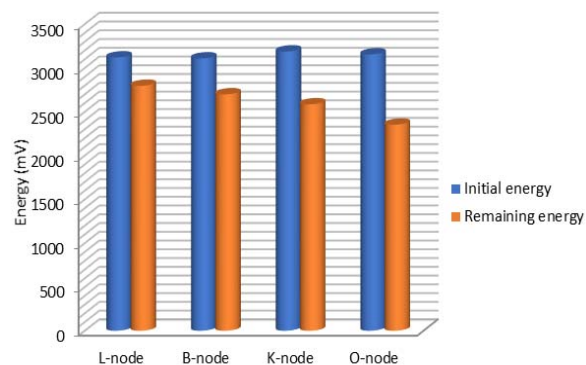

Figure 11. Power consumption for four nodes over 30 days' time

## V. CONCLUSION & FUTURE WORKS

This paper targets low cost and low power implementation for indoor air quality monitoring applications. The proposed system in this paper offers low cost design, since XBee modules and ATTiny-85 microcontrollers have been deployed, which in turn offer low cost. A sleep state algorithm and interface circuit have been designed and implemented to achieve minimum power consumption. For future works, energy efficient algorithms are required to be adopted in order to further minimize the power consumption required for indoor air quality monitoring systems.

ACKNOWLEDGMENT

The authors would like to acknowledge the financial support for this work, from the Deanship of Scientific Research (DSR), University of Tabuk, Tabuk, Saudi Arabia, under grant number S-1436-164.